\documentclass[conference]{IEEEtran}
\usepackage[draft]{common}

\IEEEoverridecommandlockouts
\usepackage{cite}
\usepackage{amsmath,amssymb,amsfonts}
\usepackage{algorithm}
\usepackage{algpseudocode}
\usepackage{graphicx}
\usepackage{textcomp}
\usepackage{listings}
\usepackage{bytefield}
\usepackage{rotating}
\usepackage{subcaption}
\usepackage{booktabs}
\usepackage[hidelinks]{hyperref}
\usepackage{numprint}
\usepackage{xfp}
\usepackage{xstring}

\newcommand{\MaticRate}{2.15}
\newcommand{\MaticToUSD}[1]{\fpeval{round(#1 * \MaticRate, 4)}}

\npdecimalsign{.}
\npthousandsep{}
\lstdefinestyle{Python}{
  morekeywords={with}
}

\makeatletter
\newcommand\addrsize{\@setfontsize\addrsize\@vipt\@viipt}
\makeatother

\newcommand{\polygonaddr}[2][\addrsize]{{#1\href{https://polygonscan.com/address/#2}{\texttt{#2}}}}
\newcommand{\polygontx}[2][\small]{{#1\href{https://polygonscan.com/tx/#2}{\texttt{\StrMid{#2}{0}{8}...\texttt{\StrMid{#2}{60}{66}}}}}}

\begin{document}

\title{Dissimilar Redundancy in DeFi}

\author{\IEEEauthorblockN{Daniel Perez}
\IEEEauthorblockA{\textit{Imperial College London} \\
}
\and
\IEEEauthorblockN{Lewis Gudgeon}
\IEEEauthorblockA{\textit{Imperial College London}} \\
}

\maketitle

\begin{abstract}
The meteoric rise of Decentralized Finance (DeFi) has been accompanied by a plethora of frequent and often financially devastating attacks on its protocols. 
There have been over 70 exploits of DeFi protocols, with the total of lost funds amounting to approximately 1.5bn USD.
In this paper, we introduce a new approach to minimizing the frequency and severity of such attacks: dissimilar redundancy for smart contracts.
In a nutshell, the idea is to implement a program logic \textit{more than once}, ideally using \textit{different} programming languages.
Then, for each implementation, the results should \textit{match} before allowing the state of the blockchain to change.
This is inspired by and has clear parallels to the field of avionics, where on account of the safety-critical environment, flight control systems typically feature multiple redundant implementations.
We argue that the high financial stakes in DeFi protocols merit a conceptually similar approach, and we provide a novel algorithm for implementing dissimilar redundancy for smart contracts.
\end{abstract}


\section{Introduction}

Decentralized Finance (DeFi) Protocol hacks are frequent - with more than 70 to date totalling losses of more than 1.5bn USD\cite{defihackscryptosec}.
For a financial infrastructure that is purportedly going to replace traditional finance, this is worrisome.
The severity of the issue of hacks is exacerbated by the non-custodial nature of DeFi systems. 
Unlike in traditional financial systems, where in the event of financial disaster, there are often safety-nets such as the state or insurers, in the DeFi setting there are no such safety provisions at scale. 
Moreover, while there is a nascent insurance market for DeFi insurance, e.g. \cite{nexusmutual}, such solutions provide only a (necessary) second-best solution: providing coverage in the event of a DeFi system failure.
A first-best solution is to prevent the failure in the first place.

Preventing such failures is challenging. 
Leaving aside the security of the underlying blockchain layer, the two main pillars of DeFi protocol security are (i) extensive smart contract testing and (ii) code audits. 
Both of these have drawbacks.
Ensuring adequate test coverage is very challenging, not least when the objective is to ensure all edge-cases are covered in the tests.
While testing is an essential part of smart contract development, we suggest that it is not realistic to expect even best practice testing with a very high degree of coverage to be sufficient to prevent \textit{all} bugs.
Code audits are often performed by external teams under time pressure, with audits typically lasting 2-3 weeks from start to finish, even for teams with no prior familiarity with the code base. 
Even for the most experienced software engineer, native to DeFi, it is wishful thinking to believe that they will always be able to catch every bug.
The problem is compounded by smart contract composability - where DeFi protocols are snapped together like DeFi lego - which serves to increase the the challenge as tests and auditors now have to anticipate bugs that could arise with as yet unseen smart contracts.

This paper presents a new approach to preventing failures at the smart contract level: dissimilar redundancy. 
In aviation, the safety-critical nature has led to the emergence of a practice of implementing multiple separate and redundant flight control systems.
For example, the Boeing 777 \cite{boing777} featured a Fly-By-Wire flight control system that had to meet extremely high levels of functional integrity and reliability.
To do this, it had three separate primary flight computers, with each computer containing three \textit{dissimilar} internal computational lanes.
The lanes differed in terms of compilers, power supply units and microprocessors, with, for example, lane 1 using the AMD 29050, lane 2 the Motorola 68040 and lane 3 the Intel 80486.
Within each of the three flight computers, two of the lanes acted as monitors while the third lane was in command.
In this way, the flight computer features a form of redundancy that is \textit{dissimilar}, with the multiple lanes being resistant to bugs induced by microprocessors or compilers.

We apply the core of this idea to smart contracts.
We implement and detail a system based on a proxy pattern which relies on dissimilar implementations of a DeFi protocol, and cross-checks one against the other before effecting any on-chain state change.
On Ethereum, this approach has already been taken for client implementations, with the community maintaining multiple open-source clients, developed by different teams and using different programming languages~\cite{ethereumclients}.
The purpose of this approach is to strengthen the network and make it more diverse, with a view to avoiding a single client dominating the network in order to remove single points of failure.
We extend this concept to the smart contract layer itself. 

Our contributions are as follows.

\begin{itemize}
    \item We introduce the notion of dissimilar redundancy for DeFi protocols
    \item We provide the first implementation of a protocol for dissimilar redundancy for a DeFi protocol\footnote{\url{https://github.com/danhper/smart-contract-dissimilar-redundancy}}
    \item We evaluate the protocol on a smart contract auction system implemented in both Solidity and Vyper, verify that a fuzzing approach would be able to detect purposefully introduced bugs, and provide the costs in USD of using a protocol for dissimilar redundancy
\end{itemize}

\subsection*{Outline}

We set out the necessary background in Section~\ref{sec:background}, provide our methodology in Section~\ref{sec:methodology}, evaluate it in Section~\ref{sec:evaluation}, consider related work in Section~\ref{sec:relatedwork} and conclude in Section~\ref{sec:conclusion}.

\section{Background}
\label{sec:background}
In this section, we provide the necessary background regarding smart contracts, their potential vulnerabilities, as well as the cost of their execution.

\subsection{Smart contracts}
Smart contracts are program objects that are native to blockchains.
In the context of Ethereum, such smart contracts require that the underlying blockchain is a transaction-based state-machine.
State changes occur on a blockchain through transactions, which are atomic: they either succeed, where the state is updated, or fail, where the state of the blockchain remains unchanged. 
For the approach we take below, atomicity is a crucial property of smart contracts as it provides that the blockchain cannot be left in an invalid or inconsistent state. 
Communication between two smart contracts occurs via message calls within the same execution context.

\subsection{Scaling solutions}
Given the high cost of using Ethereum, many scaling solutions have emerged trying to solve this issue.
At the heart of these approaches is taking computation off the layer-one blockchain, and instead performing this on a separate later while using the layer-one blockchain as an anchor of trust.
Such protocols are typically denoted layer two protocols, the name for a family of solutions that seek to allow applications to scale by processing transactions off the main blockchain.

In the context of Ethereum, one approach is to use rollups~\cite{rollups}.
Rollups perform transaction execution away from the layer-one blockchain but store transaction data on-chain.
In doing so, the security properties of layer-one are leveraged as an anchor of trust by the rollup approach, which is then able to perform transaction execution off-chain. 

For a full summary of these approaches, see~\cite{gudgeon2019sok}.

\subsection{Smart contract vulnerabilities}
Over the years, there have been many smart contract exploits often leading to significant losses of money~\cite{263870,atzei2017survey}.
There are many different ways in which smart contracts can be exploited but on a very high level, attacks can be classified as ``technical'', which means that an attacker can steal funds by exploiting a technical issue, such as a bug, in the contract, or ``economical'', which means that the attacker can exploit a contract through incentive that might not be properly aligned~\cite{DBLP:journals/corr/abs-2101-08778}.
For the purpose of this paper, we will focus on technical security of smart contracts.
Within technical security itself, there is also a myriad of different potential issues, such as contract vulnerable to re-entrancy~\cite{tubiblio111410} or flash loan attacks~\cite{qin2020attacking}.
However, another very common way in which contracts fail is simply logical bugs.
Such logical bugs have cost millions to many smart contract projects.
For example, Compound was recently victim of such an incident~\cite{compoundhack}, that cost the protocol over 90 million USD.
The error was as simple as a greater than symbol that should have been greater or equal.
\section{Methodology}
\label{sec:methodology}

\begin{figure}
  \centering
  \includegraphics[width=\columnwidth]{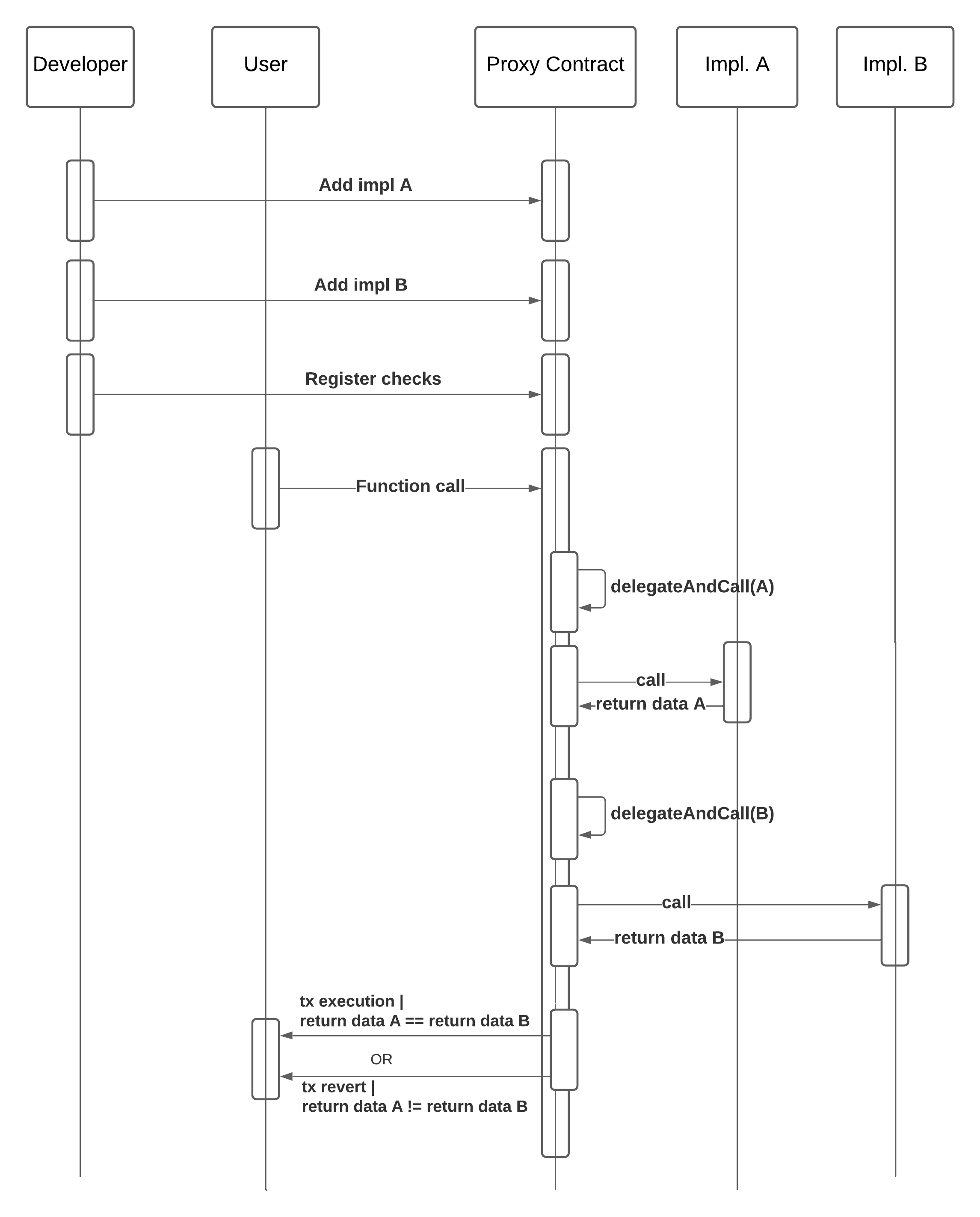}
  \caption{Overview of our dissimilar redundancy framework}\label{fig:framework-overview}
\end{figure}

\subsection{Overview}

We now turn to how we use the approach of dissimilar redundancy in the context of Ethereum.
At the centre of our approach is the use of a proxy architecture pattern.
With a proxy pattern, all message calls to a contract $\mathcal{C}$ first go through a proxy contract $\mathcal{P}$ that serves to direct the message calls to contract $\mathcal{C}$.
With this pattern, contract $\mathcal{C}$ contains the actual implementation logic while contract $\mathcal{P}$ provides a storage layer.
At present, a common use of this pattern is to provide contract upgradeability~\cite{web:eip1822,web:eip1967}: while contracts cannot be directly upgraded once deployed, upgradeability can be mimicked by changing where contract $\mathcal{P}$ delegates calls to from contract $\mathcal{C}$ to contract $\mathcal{C}_1$.

We expand on this pattern: in our pursuit of dissimilar redundancy, we allow $\mathcal{P}$ to delegate to multiple implementations at once.
This is in contrast to the standard pattern which only permits delegation to a single implementation at a time.
In a nutshell, our proxy contract $\mathcal{P}$ sequentially calls two different implementations - supposedly identical in logic - $\mathcal{C}_1$ and $\mathcal{C}_2$ and ensures that the data returned by function calls to each implementation as well as return values from an arbitrary number of \textit{checks} provided by the contract developer match.
In this context, a check is a call to a contract's function of which the return value is deemed relevant to the function called.
For example, in the context of an ERC-20 token~\cite{web:eip20}, the developer might want to add \lstinline{balanceOf(from)} and \lstinline{balanceOf(to)} as a check for \lstinline{transferFrom(from, to, amount)}.
The proxy will then call \lstinline{balanceOf} twice after each call to \lstinline{transferFrom} and ensure that the results are consistent among the implementations.

Although the overall idea is straightforward, the actual implementation requires several technical difficulties to be overcome.
We show an overview of the framework in Figure~\ref{fig:framework-overview}.

\begin{algorithm}[tb]
  \caption{Dissimilar redundancy framework call delegation}
  \label{alg:call-delegation}
  \small
  \begin{algorithmic}
    \Function{CallDelegate}{impl, data, checks, isLast}
      \State (ok, retData) $\gets$ \Call{DelegateTo}{impl, data}
      \State checkResults $\gets$ \Call{RunChecks}{checks}
      \State checksHash $\gets$ \Call{HashChecks}{checkResults}
      \If{isLast}
        \State \textbf{return} (ok, retData, checksHash)
      \Else
        \State \textbf{revert} (ok, retData, checksHash)
      \EndIf
    \EndFunction
    \Statex
    \Function{RedundantCall}{implementations, data}
      \State $n \gets$ \Call{Length}{implementations}
      \State signature $\gets$ \Call{GetSig}{data}
      \State checks $\gets$ \Call{GetChecks}{signature}
      \Statex
      \For {$i \gets 0,n-1$}
        \State impl $\gets$ implementations[$i$]
        \State last $\gets i~\text{\lstinline{==}}~n - 1$
        \Statex
        \State callData $\gets$ \Call{Encode}{CallDelegate, impl, data, checks, last}
        \State (\_, delegateRet) $\gets$ \Call{DelegateTo}{this, callData}
        \State (ok, retData, checksHash) $\gets$ \Call{Decode}{delegateRet}
        \Statex
        \If{\Call{IsDefined}{previousOk}}
          \State \Call{Assert}{previousOk \lstinline{==} ok}
          \State \Call{Assert}{previousRetData \lstinline{==} retData}
          \State \Call{Assert}{previousChecksHash \lstinline{==} checksHash}
        \Else
          \State previousOk $\gets$ ok
          \State previousRetData $\gets$ retData
          \State previousChecksHash $\gets$ checksHash
        \EndIf
      \EndFor
      \Statex
      \State \textbf{return} (ok, retData)
    \EndFunction
  \end{algorithmic}
\end{algorithm}

\subsection{The technical challenges}

\paragraph{Calling implementations with the same state}

The first difficulty is that our approach requires both implementations to be called with the same initial state, sequential contract calls would typically modify this state.
To overcome this, all state changes made by a call to an implementation need to be rolled back before we call the next implementation.
We leverage the atomic nature of Ethereum's calls to achieve this: if a transaction raises an error, the state reverts to the initial state.

In our approach, instead of the proxy directly delegating to an implementation, it first delegates to \textit{itself}, passing in the call data as an argument along with the implementation to call and the checks to perform.
In this delegated call, the proxy then delegates to the implementation, executes all the checks and combines their result in a single hash value.
It then reverts the execution to rollback the changes made by the implementation and returns the checks hash as the revert data.
The only exception to this is the call to the last implementation, where it returns the checks hash normally instead of reverting, to persist the changes.
We provide a high-level overview of the delegation logic in Algorithm~\ref{alg:call-delegation}.
Low-level implementation details can be found in our open-source implementation.

\paragraph{Check encoding}

The second difficulty concerns how the checks to be performed after each execution should be encoded.
A check is a call to a contract that needs to be consistent after each execution.
To make the proxy retain the interface of a proxied implementation, we must pre-register the checks with the proxy, rather than specifying the checks with each call.
A naive implementation would permit the developer to register a function with pre-encoded arguments to be called after any call.
However, such an implementation would have severe limitations. 
Pre-encoding arguments makes calls such as the one described above for \lstinline{balanceOf} \textit{impossible}, as these depend on call data and transaction information---this information is only available at runtime.
Since these calls need to be well targeted for them to be effective and find potential discrepancies among implementations, such an approach would not be viable.

Instead, we implement an approach which achieves the following objectives:

\begin{enumerate}
\item the registration of per-function checks, without pre-encoded arguments. For example: \lstinline{transferFrom} and \lstinline{approve} could have a different set of checks registered.
\item call data and transaction information should be accessible during the checks
\end{enumerate}

The first objective is easily achieved by storing a mapping from function signature to checks and retrieving the relevant checks depending on the function signature called by the current call to the proxy.
The second objective is more challenging, as the developer must be able to register checks upfront that rely on information only available when the function is executed.
To allow for this, rather than registering checks by passing in the arguments themselves, we design a simple byte encoding for the arguments that allows abstract arguments, registered with the checks, to be mapped to the concrete arguments that are computed when executing the checks.
For example, an abstract argument could be ``the first argument of the current call'' or ``the sender of the current transaction''.
When executing the checks, the proxy will map these to the actual value of the first argument or the sender of the transaction, and encode these when calling the check function.

\begin{figure}[tb]
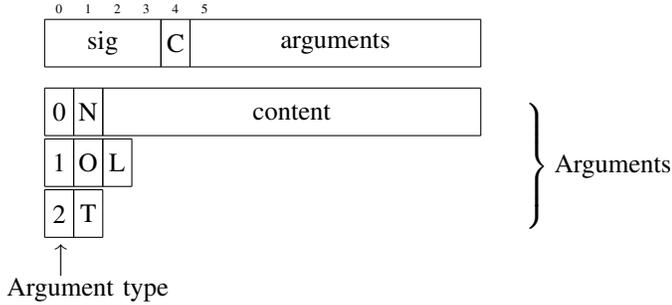

  \centering
  \begin{bytefield}[bitwidth=1.1em]{16}
    \bitheader{0-5} \\
    \bitbox{4}{sig} & \bitbox{1}{C} & \bitbox{10}{arguments}\\[2ex]
    \begin{rightwordgroup}{Arguments}
      \bitbox{1}{0} & \bitbox{1}{N} & \bitbox{13}{content}\\[0.5ex]
      \bitbox{1}{1} & \bitbox{1}{O} & \bitbox{1}{L}\\[0.5ex]
      \bitbox{1}{2} & \bitbox{1}{T}
    \end{rightwordgroup}\\
    \hspace{1mm}$\big\uparrow$\\
    \hspace{-5mm}$\text{Argument type}$
  \end{bytefield}
  \caption{Encoding for registering checks. C is the number of arguments. The three types of arguments are shown. Type 0 represents a static argument and N is the length of the static content. Type 1 represents call data argument where O and L are the offset and length to retrieve from call data. Type 2 represents environment argument and T is the type of environment variable (e.g. \lstinline{msg.sender}) to use}\label{fig:arg-encoding}
\end{figure}

In Figure~\ref{fig:arg-encoding}, we show how we encode abstract arguments to register the checks.
The first four bytes are, as for regular calls, the signature of the function to be called.
The next byte is the number of arguments to pass to the function.
Each argument can then be one of three types: a static argument, a call data argument or an environment argument.
These types determine how the concrete argument will be computed when the contract is called:

\begin{itemize}
    \item A static argument: simply passed through to the function as a concrete argument
    \item A call data argument: the given bytes from the transaction call data are extracted and passed as an argument
    \item An environment argument: looks up the concrete argument in the current transaction. 
    The argument to be looked up in the environment, e.g., the sender of the transaction or the current block timestamp is specified by a single byte in the abstract argument.
\end{itemize}

Once all the abstract arguments are converted into their concrete counterparty, they are encoded along with the function signature and the check can be performed.

\begin{figure}
\begin{lstlisting}[language=Python]
tokenProxy.registerCheck(sig["transferFrom"],
  tokenProxy, encode_args(Token, "balanceOf", [
  (ArgumentType.CallData, (4, 32))]))

tokenProxy.registerCheck(sig["transferFrom"],
  tokenProxy, encode_args(Token, "balanceOf", [
  (ArgumentType.CallData, (36, 32))]))

tokenProxy.registerCheck(sig["transferFrom"],
  tokenProxy, encode_args(Token, "allowance", [
  (ArgumentType.CallData, (4, 32)),
  (ArgumentType.Env, EnvArg.Sender)]))
\end{lstlisting}
  \caption{Example checks registration for an ERC-20 token \lstinline{transferFrom} function}\label{fig:sample-checks}
\end{figure}

This encoding provides enough flexibility to perform a wide variety of different checks.

\paragraph{Toy example}
We show an example of such checks in Figure~\ref{fig:sample-checks}, where we register three different checks for the \lstinline{transferFrom} function of an ERC-20 token.
The two first checks are for the balance of the first argument (the address sending tokens) and the second argument (the address receiving tokens) of the \lstinline{transferFrom}.
In both cases, we extract these arguments from the call data.
The third check is for the allowance and also uses the first argument but also the sender of the transaction, as their allowance is expected to decrease after a successful call to \lstinline{transferFrom}.

Now that we have established the methodology behind this approach to implementing dissimilar redundancy, we turn to an evaluation.

\section{Evaluation}
\label{sec:evaluation}
To evaluate our solution, we use a smart contract implementing a simple auction system.
The rules of the auction system are as follow:
\begin{enumerate}
\item A seller starts an auction with an NFT of their choice and sets an end time
\item The NFT is transferred to the auction contract
\item Any user can bid in the auction and the bid must be strictly greater than the previous highest until the auction ends
\item After the auction ends, anyone can ``finalize'' the auction, which will either transfer the NFT to the winner of the auction or transfer it back to the owner if there were no bids
\end{enumerate}

We implement the auction contract with the same behavior in both Solidity and Vyper but purposefully introduce a couple of implementation bugs in one of the two contracts.
We then check whether these bugs would be detected by fuzzing the contract and causing the proxy to fail due to inconsistencies in the evaluation results.

In particular, we introduce two bugs to the Vyper version of the contract.
First, rather than checking than the bid is strictly greater than the previous one, we check that the bid is greater or equal.
This means that in this particular case, the Solidity version will revert the transaction while the Vyper one will successfully execute.
Second, we omit the case where there were no bidders.
As a result, both versions will successfully execute but for the Vyper implementation, the ownership of the NFT will still be the auction contract in the case there were no bidders.

\begin{figure}
\begin{lstlisting}[language=Python]
auction_proxy.registerCheck(sig["finalize"],
  nft_collection.address,
  encode_args(
    NftCollection, "ownerOf",[
    (ArgumentType.Static, ("uint256", NFT_ID))]
))

@given(bids=st.lists(st.tuples(
   st.integers(min_value=0), addresses())))
def test_bid(bids):
  with ensure_consistent():
    for value, account in bids:
        auction_proxy.bid({"from": account,
                           "value": value})

@given(bids=st.lists(st.tuples(
   st.integers(min_value=0), addresses())))
def test_finalize(bids):
  with ensure_consistent():
    for value, account in bids:
      auction_proxy.bid({"from": account,
                         "value": value})
    chain.sleep(3600)
    auction_proxy.finalize()
\end{lstlisting}
  \caption{Testing code using dissimilar redundancy}\label{fig:testing-code}
\end{figure}

\subsection{Development-time testing}
We first test our code locally to show how our approach can make bug-detection at development time significantly easier.

To test our code, we use Python in combination with the Brownie framework~\footnote{https://eth-brownie.readthedocs.io/} for testing and the Hypothesis library~\cite{maciver2019hypothesis} to generate test cases.
We show the most important part of the code we use to our auction contract in Figure~\ref{fig:testing-code}.
We note that the \lstinline{ensure_consistent} context manager is implemented as part of our tooling and will only fail if a call reverts because implementations did not behave similarly.

In the tests, we first register a check for finalize that will look up the owner of the NFT that was on sale.
For testing both \lstinline{bid} and \lstinline{finalize}, we generate random bids, where a bid is an account and value (price to pay for the auction) pair.
For bid, we only execute all the bids while for finalize, we also ensure that the auction is ended and execute finalize.

\begin{figure}
  \begin{subfigure}{\columnwidth}
\begin{lstlisting}[basicstyle=\ttfamily\footnotesize]
Falsifying example: test_bid_consistency(
    bids=[
     (1, <Account '0x66aB6...5871'>),
     (1, <Account '0x66aB6...5871'>)],
)
===============================================
FAILED tests/test_auction.py::test_bid -
brownie.exceptions.VirtualMachineError: revert:
all implementations must return the same success
\end{lstlisting}
    \caption{Failing test case for the \lstinline{bid} function showing a failure after two bids with the same value were placed}
  \end{subfigure}
  \begin{subfigure}{\columnwidth}
\begin{lstlisting}[basicstyle=\ttfamily\footnotesize]
Falsifying example: test_finalize_consistency(
  bids=[]
)
===============================================
FAILED tests/test_auction.py::test_finalize -
brownie.exceptions.VirtualMachineError: revert:
all implementations must return the same checks
\end{lstlisting}
    \caption{Failing test case for the \lstinline{finalize} function showing a failure when no bids have been placed}
  \end{subfigure}
  \caption{Failing tests when fuzzing the auction contract with a correct and an incorrect implementation}\label{fig:test-results}
\end{figure}

Using this approach, trying to fuzz the contract requires performing differential fuzzing on the different implementations of the contract logic registered by the proxy.
This makes it possible to easily identify the cases where the two implementations do not behave in the same way.

In Figure~\ref{fig:test-results}, we show the results of these tests.
Note that we fix the first failing test before proceeding to the second one.

For the \lstinline{bid} function, the test framework correctly outputs a failure when two bids are placed with the same value in a row.
The failure scenario is clear from the test output, as the two bids of the input both have a value of 1.
The error message mentions that all implementations should return the same ``success'', which means that one of the implementations successfully executed (the bug-ridden version), while the other failed to executed.

The test for the \lstinline{finalize} function also correctly fails with an example containing no bids.
This is consistent with the bug in the Vyper version of the Solidity which does not transfer back the token properly to its original owner.
The failing test also mentions that all implementations must return the same checks which means that all implementations had the same success status (they all succeeded in this particular case) but the checks did not return the same value.
Indeed, a check was registered to look up the NFT owner.

Overall, with this example, we have seen that with only a few lines of code, it was possible to have extensive coverage of the tested function that is able to automatically find discrepancies among implementations and indicate to the developer the test cases that would yield different results.

\begin{table}[ht]
  \setlength{\tabcolsep}{3pt}
  \begin{tabular}{l r}
    \toprule
    \textbf{Name} & \multicolumn{1}{l}{\textbf{Address}}\\
    \midrule
    Auction proxy & \polygonaddr{0xAd837BDD116C14aA82311Db7D1879C7cDDCfd283}\\
    Auction (Sol, proxied) & \polygonaddr{0xdb85f3DB2aA6E5e294485972ABE921be188b6A37}\\
    Auction (Vy, proxied) & \polygonaddr{0x9FD31161360B5E772f2b9C469D4A35E679273Dbf}\\
    Auction (Sol, standalone) & \polygonaddr{0xE575CCb0213393eBFc9258013af1c43e9E416544}\\
    Auction (Vy, standalone) & \polygonaddr{0xEe164319fE07127Efc8fdf8b3e99ea736F8c955E}\\
    \bottomrule
  \end{tabular}
  \caption{Contracts deployed on Polygon mainnet. ``Sol'' are Solidity contracts. ``Vy'' are Vyper contracts. ``Proxied'' are contracts used by the proxy. ``standalone'' are contracts interacted with directly.}\label{tab:deployed-contracts}
\end{table}

\subsection{Real-world deployment}
An important strength of our approach is that it is possible to utilize the two implementations not only at development and testing time but also after the contract is deployed, ensuring that all the transactions executed will always be consistent across the different implementations provided.
To demonstrate how this would perform on a real-world blockchain, we deploy our auction and its two implementations on the Polygon main network, ensure that the calls that would trigger revert correctly, and measure the cost overhead of our approach.
We provide a list of all the deployed contracts in Table~\ref{tab:deployed-contracts}.

To be able to give comparable results, we deploy our proxy using the Solidity and Vyper implementations, as well as a standalone version of each implementation.
During our interactions with the contracts, we maintain a fixed gas price of 30 Gwei, which at the time of writing was enough for near instantaneous inclusion in a Polygon block.

\begin{table}[tb]
  \setlength{\tabcolsep}{10pt}
  \begin{tabular}{l r r r r}
    \toprule
    & Start & First bid & Subsequent bid & Finalize\\
    \midrule
    Proxied & \MaticToUSD{0.0106683} & \MaticToUSD{0.00428409} & \MaticToUSD{0.00280368} & \MaticToUSD{0.00672003}\\
    Solidity & \MaticToUSD{0.00436455} & \MaticToUSD{0.002097} & \MaticToUSD{0.00135948} & \MaticToUSD{0.00242331}\\
    Vyper &  \MaticToUSD{0.0050136} & \MaticToUSD{0.00209271} & \MaticToUSD{0.00134982} & \MaticToUSD{0.003282}\\
    \bottomrule
  \end{tabular}
  \caption[Function cost]{USD cost\footnotemark~of calling different functions of the auction contract with and without dissimilar redundancy proxy. The gas price is fixed to 30 Gwei.}\label{tab:tx-costs}
\end{table}

\footnotetext{We use the December 12th 2021 price of 2.15 USD per MATIC token, Polygon's native token used to pay for gas fees}

We summarize the cost nominated in US dollars of all the interactions with our auction contracts in Table~\ref{tab:tx-costs}.
We split the costs of the first and the subsequent bids, since the first bid allocates storage, using more gas than subsequent ones.
Since the proxy is using both the Solidity and the Vyper implementation, a lower bound for the cost is the sum of the cost of each individual implementation.
The difference between the sum of these costs and the cost of calling the proxy is the overhead of the proxy itself, including the cost of calling checks after each call and checking consistency among results.
In our example, only \lstinline{finalize} has a check registered, to check for the owner of the NFT after execution.
For the calls to \lstinline{start} and \lstinline{bid}, respectively about 1.2\% and 2.2\% of the total cost is part of the proxy overhead, the rest of the cost being used by the actual underlying functions.
For \lstinline{finalize}, the overhead is understandably higher as a check is registered.
Indeed, about 14.5\% of the cost is used by the proxy itself.

We also check that a transaction that would cause our correct and our buggy implementation to diverge would correctly revert.
To do so, we bid using subsequently twice the same amount, which fails for the proxied version (tx \polygontx{0x5f840ae129200b9d908714081e59697bc03c6cc2a34460b1d9b9002616dc334c}) by returning the same error as the one we saw during the tests.
As a sanity check, we try the same sequence of bids on the standalone implementations and the Solidity version reverts correctly (tx \polygontx{0x406ad7356e54cd539b7e1820b479df9346e56bd7534373dd809c52f127759673}) while the Vyper one succeeds (tx \polygontx{0x888189e5681f34ce776244bcf6eedb911760b0e1420c6df6053c96c387df41e8}).

\section{Limitations}
\label{sec:limitations}
While our approach comes with strong benefits in terms of reliability, it has some limitations.
In this section, we will go over these limitations and provide details on how some of them could be tackled.

\point{Transaction fees} As we have seen in the previous section, this approach will always at least double the cost of transactions and potentially increase it further if many calls need to be performed.
There is not any direct way to prevent this issue or improve it significantly at the implementation level.
This means that using such an approach on a expensive network, such as Ethereum, is almost unfeasible, as the price increase for transactions would likely be unacceptable for many users.
However, more and more layer two solutions are being developed, with transaction fees orders of magnitude lower than what can be seen on Ethereum mainnet.
This is for example the case of Polygon mainnet, which we used for the evaluation earlier.
As we can see in Table~\ref{tab:tx-costs}, although the costs have been doubled, this represents an increase in the order of a cent in the worst case.
As Ethereum is moving towards a layer 2 future~\cite{eth-endgame} and transaction fees become negligible, the fee overhead of our approach might become less and less of a concern.

\point{Development cost} Another obvious limitation is the development cost of another implementation of the same contract.
However, there are several arguments why this might not be a critical issue.

First, such an approach has been seen with Ethereum clients, which are vastly larger in terms of complexity and code base than most protocols built on top of smart contracts.
In a similar way that the Ethereum Foundation distributes fund to help external teams maintain clients, a well established protocol could potentially operate similarly to have alternative implementation maintained.

Second, given the often enormous amount of money at stake, a significant part of the smart contract development goes into testing rather than simply implementing the contract.
For example, Uniswap V3~\cite{web:uniswap} has almost twice more test code than actual implementation.
Our approach being highly complementary to regular testing, it could be integrated as part of the development process as yet another way to reduce the number of potential bugs or vulnerabilities in the contract.

\point{Storage layout} Another limitation of our approach is that all the implementations must use exactly the same storage layout.
The proxy contract will be storing all the state of the contract, so all implementations must read at the same location in storage to be able to retrieve the information they are looking for.
One of the main drawback is that it imposes some rigidity on alternative implementations, which might make it less likely to try implementing the logic in very similar way and potentially replicate bugs in multiple implementations.
Another issue is that this makes it hard to combine implementations written in Solidity and Vyper.
The two languages use mostly the same mapping from state variable to Ethereum storage slot but for mappings, they use a very slightly different way to compute the storage slot which makes it impossible to use this approach for any contract using mappings.
However, this issue is trivial to fix, as it would only require a very minor, albeit backward incompatible, change in the Vyper compiler.

\section{Related work}
\label{sec:relatedwork}
We first present how a similar approach has been helpful for Ethereum clients and then present some related research discussing differential fuzzing.

\point{Ethereum clients} From early on, the Ethereum network has been running using different client implementations.
The goal of having multiple clients has always been to make the network more robust and ensure that it does not fail in case there would be a bug, a vulnerability, or an avenue for a potential denial-of-service (DoS) attack in one of the implementations.
In the case of Ethereum, this allows for several failure modes.
If one of the implementations had a bug that would make it crash on certain transactions, the network would continue to operate with other implementations that do not contain this bug.
This would be similar in the case of a DoS attack only effective on one type of implementation.
On the other hand, if a bug or exploit would result in a different state transition after executing a transaction, it would create a network split where the victim implementation would only manage to reach consensus with nodes using the same implementation.
This is riskier than the previous case but remains safe as long as exchanges and other entities bridging on-chain activity to off-chain assets ensure all implementations are in a consistent state before accepting to process a transaction.
Overall, this diversity of clients has been very beneficial to Ethereum, despite the high maintenance cost, and has allowed it to operate smoothly for over 6 years.

\point{Differential fuzzing} Having two implementations that should behave in the same way allows to perform differential fuzzing: fuzzing both implementations trying to look for cases where they would behave differently.
This technique has already been used in multiple domains such as cryptography~\cite{brubaker2014using}, programming languages~\cite{chen2016coverage}, and blockchain consensus~\cite{evmlab,fu2019evmfuzzer}.
A recent work leveraging differential fuzzing to find bug in Ethereum clients~\cite{273753} has managed to find not only most known consensus bugs but also two new ones, including a bug that led to a fork in the consensus due to only part of the full nodes in the network having upgraded to the latest version~\cite{geth-fork-bug}.
Overall, this shows the potential of differential fuzzing and how it can be useful for finding bugs and zero-day exploits.

\section{Conclusion}
\label{sec:conclusion}
We have argued that the high financial stakes in the context of DeFi merit an approach to program redundancy inspired by avionics: the utilization of dissimilar redundancy. 
Through implementing the same program logic more than once, ideally with different programming languages and even by different engineering teams, and then using an on-chain execution logic that ensures that the dissimilar implementations must \textit{agree} before the on-chain state can update, redundancy is brought into the smart contract ecosystem.
Such redundancy should serve to make smart contracts, and DeFi as a whole, less vulnerable to exploits from implementation bugs.

We hope that this paper can offer one step on the path to a more robust and secure DeFi.
As in avionics, in DeFi, the stakes are high, and the risks real.
\bibliographystyle{IEEEtran}
\bibliography{references}



\end{document}